\documentclass[11pt, letterpaper]{article}

\usepackage{graphics}
\usepackage{graphicx}
\usepackage{epsfig}
\usepackage{amssymb}
\usepackage{natbib}
\usepackage{aas_macros}
\usepackage{multicol}
\usepackage{color}

\makeatletter
\usepackage{charter}

\usepackage{amsmath}
\usepackage{subfigure}
\usepackage{multirow} 
\usepackage{amsmath,amssymb,mathrsfs}
\usepackage{hyperref} 
\usepackage{ulem}

\usepackage{aas_macros}

\usepackage[usenames,dvipsnames]{xcolor}

\newcommand{\cm}{\textcolor{black}} 
\newcommand{\rvh}{\textcolor{black}} 

\title{\vspace{-15mm}\fontsize{24pt}{10pt}\selectfont\textbf{Analysis of the
first IPTA Mock Data Challenge by the EPTA timing data analysis working group}} 

\author{
  \large
  \textsc{Rutger van Haasteren} \thanks{vhaasteren@gmail.com}\\[1mm] 
  \normalsize Max-Planck-Institut f{\"u}r Gravitationsphysik (Albert-Einstein-Institut), D-30167 Hannover, Germany
  \vspace{5mm} \\
  \large
  \textsc{Chiara~M.~F. Mingarelli and Alberto Vecchio}\\[1mm]
  \normalsize School of Physics and Astronomy, University of Birmingham, Edgbaston, Birmingham B15 2TT, UK 
  \vspace{5mm} \\
  \large
  \textsc{Antoine Lassus}\\[1mm] 
  \normalsize LPC2E,CNRS, Universit{\'e} d'Orl{\'e}ans \\
  \vspace{-5mm}
}

\begin{document}

%
 

\maketitle

  \begin{abstract}
    This is a summary of the methods we used to analyse the first IPTA Mock Data
    Challenge (MDC), and the obtained results. We have used a Bayesian analysis in the
    time domain, accelerated using the recently developed ABC-method which
    consists of a form of lossy linear data compression.  The TOAs were first
    processed with {\rm Tempo2}, where the design matrix was extracted for use
    in a subsequent Bayesian analysis. We used different noise models to
    analyse the datasets: no red noise, red noise the same for all pulsars, and
    individual red noise per pulsar. We sampled from the likelihood with four
    different samplers: ``emcee'', ``t-walk'', ``Metropolis-Hastings'', and
    ``pyMultiNest''. All but emcee agreed on the final result, with emcee
    failing due to artefacts of the high-dimensionality of the problem. An
    interesting issue we ran into was that the prior of all the 36 (red) noise
    amplitudes strongly affects the results. A flat prior in the noise amplitude
    biases the inferred GWB amplitude, whereas a flat prior in log-amplitude
    seems to work well. This issue is only apparent when using a noise model
    with individually modelled red noise for all pulsars.
    Our results for the blind challenges are in good agreement with the injected
    values. \rvh{For the GWB amplitudes we found $h_c = 1.03 \pm 0.11
    [\times 10^{-14}]$, $h_c = 5.70 \pm 0.35  [\times 10^{-14}]$, and $h_c =
    6.91 \pm 1.72  [\times 10^{-15}]$, and for the GWB spectral index we found
    $\gamma = 4.28 \pm 0.20$, $\gamma = 4.35 \pm 0.09$, and $\gamma = 3.75 \pm
    0.40$. We note that for closed challenge $3$ there was quite some covariance
    between the signal and the red noise:
    if we constrain the GWB spectral index
    to the usual choice of $\gamma = 13/3$,} we obtain the estimates: $h_c = 10.0
    \pm 0.64 [10^{-15}]$, $h_c = 56.3 \pm 2.42 [10^{-15}]$, and $h_c = 4.83 \pm 0.50
    [10^{-15}]$, with one-sided $2\sigma$ upper-limits of: $h_c \leq 10.98
    [10^{-15}]$, $h_c \leq 60.29 [10^{-15}]$, and $h_c \leq 5.65 [10^{-15}]$.
  \end{abstract}



  \section{Introduction}
    We describe our attempts to analyse the International Pulsar Timing Array
    (IPTA) first Mock Data Challenge (MDC). Our analysis methods are implemented
    in the European Pulsar Timing Array (EPTA) data analysis library, an open
    source universal pulsar timing data analysis library under construction soon
    to be released to the public. This library is written in a combination of
    Python and C, and can interface with the pulsar timing package {\rm Tempo2}
    \citep{Hobbs2006}. 

    In Section~\ref{sec:ptaan} we briefly review the different Bayesian analysis
    techniques that we have employed. In Section~\ref{sec:setup}, we discuss the model and the setup of the
    analysis  with the results presented in Section~\ref{sec:analysis}. We conclude with some remarks in
    Section~\ref{sec:conclusions}.

  \section{PTA data analysis} \label{sec:ptaan}
    We use the Bayesian analysis method first outlined in
    \citet{vanhaasteren2009}, and further developed in
    \citet[][hereafter vHL]{vanhaasteren2010,
    vanhaasteren2011, vanhaasteren2012a}. A substantial reduction of the 
    processing time
    is obtained by applying the ABC-method of
    \citet{vanhaasteren2012b} to the data. We briefly review the likelihood function we
    use in Section~\ref{sec:review}, and we outline the ABC-method in
    Section~\ref{sec:abc}.

    \subsection{Review of the likelihood} \label{sec:review}
      We consider $k$ pulsars, with $n^{\prime}_{a}$ TOAs for the $a$-th pulsar,
      where the $n^{\prime}=\sum_{a=1}^{k}n^{\prime}_{a}$ TOAs are described as
      an addition of a
      deterministic and a stochastic part. In the observations this distinction
      is blurred because we cannot
      fully separate the stochastic contributions from the deterministic
      contributions. In practice we therefore work with timing
      residuals that are produced using first estimates $\beta_{0i}$ of the
      $m$ timing-model parameters $\beta_i$ ($i$ between $1$ and $m$); this
      initial guess is usually assumed to be accurate enough to use a linear
      approximation of the
      timing-model \citep{Edwards2006}. Here $m=\sum_{a=1}^{k}m_{a}$ is the sum
      of the number of timing-model parameters of all the individual pulsars.
      In this linear approximation, the
      timing-residuals depend on $\xi_i = \beta_i - \beta_{0i}$ as:
      \begin{equation}
	\vec{\delta t}^{\prime} = \vec{\delta t}^{\rm prf} + M\vec{\xi},
	\label{eq:prefitresidual}
      \end{equation}
      where $\vec{\delta t}^{\prime}$ are the timing-residuals in the linear
      approximation to the timing-model, $\vec{\delta t}^{\rm prf}$ is the
      vector of pre-fit timing-residuals, $\vec{\xi}$ is the vector with
      timing-model parameters for all $k$ pulsars, and the $(n^{\prime}\times m)$ matrix
      $M$ is the so-called design matrix
      \citep[see e.g. \S $15.4$ of][vHLML]{Press1992}, which describes how the
      timing-residuals depend on the model parameters. Take for example  a
      simple timing model which only contains quadratic spindown, the matrix $M$
      is a $(n^{\prime}\times 3)$ matrix, with the $j$-th column describing a
      $(j-1)$-th order polynomial. The elements of $M$ are then:
      $t_{i}^{j-1}$, with $t_{i}$ the $i$-th TOA.

      In their search for a simplified representation of the analytic
      marginalisation procedure, vHL decomposed the design matrix into an
      orthogonal basis based on the singular value decomposition 
      $M = U \Sigma V^{*}$, where $U$ and $V$ are $(n^{\prime}\times n^{\prime})$ and $(m\times
      m)$ orthogonal matrices, and $\Sigma$ is an $(n^{\prime}\times m)$ diagonal matrix.
      The first $m$ columns of $U$ span the column space of $M$, and the last
      $n=n^{\prime}-m$ columns of $U$ span the complement. We denote these two subspace
      bases as $F$ and $G$ respectively: $U = \begin{array}{cc}(F & G)\end{array}$.
      vHL showed that the likelihood can be rewritten as:
      \begin{equation}
	\label{eq:marginalisedlikelihoodnew}
	\int \! \mathrm{d}^{m}\vec{\xi} P(\vec{\delta t}^{\prime} | \vec{\xi}, \vec{\phi})
	= \frac{
	 \exp\left(-\frac{1}{2} \vec{\delta
	 t}^{\prime T}G \left(G^{T}C^{\prime}G\right)^{-1}G^{T}
	 \vec{\delta t}^{\prime}\right)
	}{\sqrt{(2\pi)^{n}\det
	\left(G^{T}C^{\prime}G\right)}}\cm{.}
      \end{equation}

    \subsection{The ABC-method} \label{sec:abc}
      \citet{vanhaasteren2012b} introduced a linear transformation called
      ``the ABC-method'' that reduces the dimensionality of the
      dataset without losing a significant amount of information from a signal
      of interest. The ABC-method is both a marginalisation over all timing
      model parameters and a lossy linear data compression method.
      
      The transformation $\vec{x} = H\vec{\delta t}$, with $\vec{x}$ the
      compressed data called the ``generalised residuals''. The likelihood of the
      generalised residuals is:
      \begin{equation}
        \label{eq:compressedlikelihood}
	P(\vec{x} | \vec{\phi})
	= \frac{
	 \exp\left(-\frac{1}{2} \vec{x}^{T} \left(H^{T}CH\right)^{-1}
	\vec{x}\right)
	}{\sqrt{(2\pi)^{l}\det\left(\Sigma_w \right)\det
	\left(H^{T}CH\right)}}.
      \end{equation}
      Although Equation (\ref{eq:compressedlikelihood}) is valid for any $H$, we
      would like to choose $H$ such that no information about the GWB signal is
      lost. This is accomplished by taking $H =
      \Sigma_w^{-1/2}W$, with $\Sigma_w$ a conservative estimate of the noise
      (consisting of the TOA uncertainties), and the columns of the $(n\times
      l)$ matrix $W$ consist of the first $l$ eigenvectors of the matrix $C^{w} =
      \Sigma_w^{-1/2}S\Sigma_w^{-1/2}$, with $S$ the covariance matrix of the
      expected signal.

      The ABC-method can be accelerated further by using an interpolation scheme
      for the compressed covariance matrix. We use the method outlined in van
      \citet{vanhaasteren2012b}, which combines linear data compression with a
      cubic spline interpolation technique.
      We divided the interval $1.04 < \gamma < 6.99$ in $118$ sub-intervals,
      where on each interval the elements of the compressed covariance matrix
      are approximated by cubic functions that are matched in value and
      derivative at the boundaries.

  \section{Setup of the analysis} \label{sec:setup}
    The basis of our analysis is the likelihood function
    described by Equation~(\ref{eq:compressedlikelihood}). In this section we discuss the
    approximations we have made in using this expression as our likelihood, and
    we discuss the model we have used to analyse the closed sets.
    
    \subsection{Justification of the model}
      \cm{The following are approximations we have made in the derivation of the likelihood of
      Equation~(\ref{eq:compressedlikelihood}):}
      
      \begin{enumerate}
	\item
	  The timing model has been linearised, which is equivalent to using a
	  Fisher-matrix approximation at the maximum likelihood value of the
	  timing model parameters \citep{Edwards2006}.
	\item
	  The data has been compressed with the ABC-method, which by design throws
	  away some information.
	\item
	  The noise and GWB are assumed to be Gaussian processes.
      \end{enumerate}
      \cm{With regard to item 1, \cite{Taylor2012} have found that the timing solutions of several pulsars
      in open challenge $2$ and closed challenge $2$ had not yet converged. This invalidates the
      maximum likelihood expansion, as some pulsars may have appeared to be noisier than they should have been.}
      \cm{Item 2 is an approximation which we have full control over, as we}
      use the ABC-method with a fidelity $\mathscr{F} \ge 0.99$ to not throw
      away too much information. \cm{Item 3 is justified by the
      central limit theorem.}

    \subsection{Model for the data challenge sets}
      We have modelled all the closed sets of the IPTA Mock Data challenge the
      same way. In our likelihood, we included:
      \begin{itemize}
	\item
	  Post-fit timing residuals as reported by {\rm Tempo2} with
	  ``-residuals''
	\item
	  TOA uncertainties as reported by {\rm Tempo2}. No variable constant
	  multiplicative \sout{fudge} factor (``EFAC'') was included.
	\item
	  A Fisher-matrix approximation to the timing model was used, as
	  obtained through the {\rm Tempo2} plugin ``-designmatrix''.
	\item
	  A red timing noise contribution for each pulsar was assumed, with a
	  power-law power spectral density as described in vHLML. The
	  low-frequency cut-off was taken to be $1/20 {\rm yr}^{-1}$; the
	  likelihood of the compressed data,
	  Equation~(\ref{eq:compressedlikelihood}) is insensitive to this value.
	  There are therefore only two free parameters for such a noise source:
	  the amplitude A, and the spectral index $\gamma$.
	\item
	  An isotropic stochastic background of gravitational waves, correlated
	  according to the Hellings \& Downs correlation curve
	  \citep{Hellings1983} was assumed, with a power-law spectral density.
	  As for the red timing noise, this signal had two free parameters in
	  our model: the amplitude and the spectral index.
	\item
	  The priors were uniform for all the spectral index parameters. The
	  spectral indices for both the timing noise and the GWB were bound to
	  the interval $\gamma \in (1,7)$.
	\item
	  The priors were uniform in $\log A$ for all amplitude parameters
	  ($1/A$ when sampling in $A$), for both the GWB and the timing
	  noise.  In units of the GWB, the amplitude parameters were all
	  assumed to lie in the range $A \in (10^{-16}, 10^{-12})$.
      \end{itemize}

      With this way of modelling we had $2\times 36 + 2=74$ free parameters in
      our model, plus over $200$ timing model parameters that we analytically
      marginalised over.

      We have also used models in which the timing noise was the same for all pulsars. This reduces the number of parameters to $2+2=4$, which makes even
      {\rm MultiNest} \citep{Feroz2009} able to sample the posterior with ease. However, we
      believe that such a model is unphysical, and we choose to list the $74$
      parameter model as our final answers.

    \subsection{Choice of prior}
    \rvh{
      Tests on the open challenge sets indicate that the prior for the red
      timing noise amplitude and the GWB amplitude needs to be uniform on a
      logarithmic scale. A prior uniform in the noise amplitude seems to
      strongly bias the results. Although it requires a more theoretical
      explanation, which we will address in a future work, a straightforward
      explanation can be given in terms of Jeffreys prior. All open and closed
      challenges have a negligible amount of timing noise in all pulsars, which
      makes the problem of estimating the timing noise amplitude equivalent to
      estimating the variance in a zero-mean Gaussian distributed time series.
    In such a problem, Jeffreys prior is uniform on a logarithmic scale.}

      \rvh{For all challenges, using a uniform prior in the noise amplitude strongly
      biases the GWB amplitude to about $75\%$ of its true value, in the case
      the timing noise was modelled for each pulsar individually. In the case
      the timing noise in all pulsars is modelled by one common parameter, this
      dependence on the prior is not present.
    }

    \subsection{The data compression setup}
      The data compression was set up as outlined in \citet{vanhaasteren2012b}.
      We used the TOA uncertainties, as
      obtained from {\rm Tempo2}, as estimates for the timing noise. The GWB signal estimate used in the ABC
      compression was  $h_c=10^{-14}$,
      $h_c=6\times 10^{-14}$, and $h_c=10^{-14}$  for the three blind challenges, respectively. These are rough estimates
      obtained through Equation~($24$) of \citet{vanhaasteren2012a}. In applying
      that equation, we left out J$0437$-$4714$ in closed dataset $2$, as this
      pulsar seems to contain a lot of high-frequency noise.

    \subsection{The different samplers}
      \subsubsection{twalk}
	We initially used the twalk \citep{Christen2010} to sample our
	$74$-dimensional parameter space. The twalk is a Markov Chain Monte
	Carlo (MCMC) sampler that has one convenient property: it does not need
	to be tuned for most distributions. This is achieved by using four
	different types of steps, one of which is the ``stretch move'' also used
	by the recently introduced and much-celebrated sampler ``emcee''
	\citep{Foreman2012}. However, twalk is not efficient in many
	dimensions, since it updates parameters with only a few dimensions at a
	time. 

	We ran the twalk for an initial burn-in of $5\times 10^{4}$ steps on all
	datasets. In this burn-in chain we judged whether or not the chain had
	come to equilibrium by looking at the log-likelihood values of the
	chains. We cut off the begin of the chain, and determined the rms
	variations of all parameters in the rest of the chain.

      \subsubsection{Metropolis-Hastings}
	Metropolis-Hastings is quite an efficient sampler, provided that the
	proposal distributions are tuned well for the problem. The acceptance
	fraction should approximate $25\%$ \citep{Roberts1997}, and mixing
	should be similar for all parameters. Since we have $74$ dimensions,
	this is not trivial to get right. We therefore chose to use the results
	of the twalk algorithm to tune the proposal distribution for a
	subsequent Metropolis-Hasgings run.
	We used a Gaussian proposal
	distribution, centered around the current sample, with a width equal to
	the parameter rms as determined from the twalk burn-in chain. We found
	that we needed to multiply all proposal widths with a factor of about $a
	\approx 0.05$ to obtain an acceptance fraction of $25\%$.

	We ran the Metropolis-Hastings algorithm for $2\times 10^{5}$ steps. The
	autocorrelation of the chain was a few $10$s of samples in this chain,
	whereas for the twalk it was over $1000$ samples.

      \subsubsection{Emcee}
        Goodman and Weare introduced an affine invariant sampler with several
	convenient properties: fast mixing, only one tunable parameter,
	invariance to all affine (linear) transformations. This algorithm has
	been implemented in the Python package ``emcee'' \citep{Foreman2012}. 

	We found that for this particular problem emcee did not
	perform very well. Many proposed steps were actually rejected because
	they fell outside of the prior range we set on the spectral indices.
	Using a sigmoid transformation (arctangent) to remove the hard
	boundaries on parameter space, effectively having emcee sampling on an
	unbounded parameter space, did not help: the acceptance fraction
	remained around $5\%$. This low acceptance fraction prohibited the chain
	from mixing properly, and we found a well-tuned Metropolis-Hastings
	algorithm to be more efficient.

      \subsubsection{\rm MultiNest}
	{\rm MultiNest} \citet{Feroz2009} is a Bayesian inference method, based on
	the ideas of nested sampling (Skilling 2000). {\rm MultiNest} uses a
	form of rejection sampling: it repeatedly draws samples from the prior
	distribution, under the restriction that the likelihood values are above
	a certain threshold that increases during the run of the analysis.

	We found {\rm MultiNest} to be slow on these $74$-dimensional parameter
	spaces.  Even though the likelihood appears to be uni-modal, most of the
	proposed samples ended up being rejected near the end of a run. We have
	tried several different choices of live points: $200$, $500$, $1000$,
	and $2000$, all with similar results. Efficiency was set to $0.99$. The
	number of samples required to get reasonable credible regions (no
	evidence convergence yet) was around $50$M points. This is approximately
	$10^3$ times more samples than was required to have good looking
	credible regions with Metropolis-Hastings. However, {\rm MultiNest} did
	provide an independent check of convergence of our MCMC chains.
	For our four-dimensional searches {\rm MultiNest} performance was closer
	to the MCMC methods.

  \section{Analysis} \label{sec:analysis}
    \subsection{The ABC-method results}
      The level of compression depends on the signal to noise ratio. We use the
      conservative approach outlined in \citet{vanhaasteren2012b}: only use the
      TOA uncertainties and an (over)estimate of the signal strength. The signal
      amplitudes we assumed for the three closed datasets were:
      $h_c=10^{-14}$, $h_c=6\times 10^{-14}$, $h_c=10\times 10^{-14}$. This
      resulted in a number of generalised residuals that was different for each
      pulsar in each dataset, because the timing model and the TOA uncertainty
      differs per pulsar. The result is presented in
      Table~\ref{tab:compression}.

      \begin{table}[ht]
	\label{tab:compression}
	\begin{minipage}[b]{0.50\linewidth}
	  \centering
	  \begin{tabular}{| l | r | r | r |}
	    \hline
	    {\bf Pulsar} & Closed 1 & Closed 2 & Closed 3 \\
	    \hline
	    \hline
	    J0030+0451 & 10 & 11 & 6 \\
	    \hline
	    J0218+4232 & 11 & 4 & 2 \\
	    \hline
	    J0437-4715 & 10 & 41 & 19 \\
	    \hline
	    J0613-0200 & 10 & 9 & 5 \\
	    \hline
	    J0621+1002 & 10 & 3 & 2 \\
	    \hline
	    J0711-6830 & 10 & 6 & 4 \\
	    \hline
	    J0751+1807 & 11 & 8 & 5 \\
	    \hline
	    J0900-3144 & 11 & 6 & 3 \\
	    \hline
	    J1012+5307 & 10 & 11 & 6 \\
	    \hline
	    J1022+1001 & 11 & 12 & 6 \\
	    \hline
	    J1024-0719 & 10 & 13 & 6 \\
	    \hline
	    J1045-4509 & 10 & 5 & 3 \\
	    \hline
	    J1455-3330 & 10 & 6 & 3 \\
	    \hline
	    J1600-3053 & 5 & 13 & 6 \\
	    \hline
	    J1603-7202 & 10 & 7 & 5 \\
	    \hline
	    J1640+2224 & 9 & 13 & 6 \\
	    \hline
	    J1643-1224 & 9 & 8 & 5 \\
	    \hline
	    J1713+0747 & 10 & 35 & 17 \\
	    \hline
	  \end{tabular}
	\end{minipage}
	\begin{minipage}[b]{0.50\linewidth}
	  \centering
	  \begin{tabular}{| l | r | r | r |}
	    \hline
	    {\bf Pulsar} & Closed 1 & Closed 2 & Closed 3 \\
	    \hline
	    \hline
	    J1730-2304 & 10 & 7 & 4 \\
	    \hline
	    J1732-5049 & 10 & 6 & 3 \\
	    \hline
	    J1738+0333 & 9 & 13 & 6 \\
	    \hline
	    J1741+1351 & 12 & 17 & 8 \\
	    \hline
	    J1744-1134 & 10 & 18 & 8 \\
	    \hline
	    J1751-2857 & 11 & 7 & 4 \\
	    \hline
	    J1853+1303 & 11 & 17 & 8 \\
	    \hline
	    J1857+0943 & 10 & 12 & 6 \\
	    \hline
	    J1909-3744 & 10 & 35 & 16 \\
	    \hline
	    J1910+1256 & 11 & 17 & 8 \\
	    \hline
	    J1918-0642 & 10 & 6 & 4 \\
	    \hline
	    J1939+2134 & 10 & 51 & 25 \\
	    \hline
	    J1955+2908 & 10 & 15 & 7 \\
	    \hline
	    J2019+2425 & 10 & 7 & 5 \\
	    \hline
	    J2124-3358 & 10 & 6 & 3 \\
	    \hline
	    J2129-5721 & 10 & 6 & 4 \\
	    \hline
	    J2145-0750 & 6 & 9 & 6 \\
	    \hline
	    J2317+1439 & 10 & 13 & 6 \\
	    \hline
	  \end{tabular}
	\end{minipage}
	\caption{The number of generalised residuals per pulsar after
	  compression for the closed datasets.  }
      \end{table}

    \subsection{Results}
      Our results for the GWB in the $74$-dimensional searches are listed in
      Table~\ref{tab:results}.
      We have calculated the errors as being the half-width of the $68\%$
      probability density region.

    \begin{table}
      \label{tab:results}
      \centering
      \begin{tabular}{| l | l | l | l | l |}
	\hline
	{\bf Dataset} & {\bf $h_c$ ($10^{-15}$)} & true $h_c$ & {\bf $\gamma$} &
	true {\bf $\gamma$} \\
	\hline
	\hline
	{\it Closed 1} & $10.31 \pm 1.09$ & $10$ & $4.28 \pm 0.20$ & $4.33$ \\
	\hline
	{\it Closed 2} & $57.03 \pm 3.48$ & $60$ & $4.35 \pm 0.09$ & $4.33$ \\
	\hline
	{\it Closed 3} & $6.91 \pm 1.72$ & $5.0$ & $3.75 \pm 0.40$ & $4.33$ \\
	\hline
      \end{tabular}
      \caption{\rvh{The inferred parameters of the GWB signal of the closed
      challenges.}}
    \end{table}

      We note that the spectral index of closed $3$ has a highly non-Gaussian
      shape. If we would list the $95\%$ boundaries instead of the $68\%$ we
      would obtain: $\gamma = 3.97 \pm 0.89$, which makes the true value of $4.33$
      well within our $95\%$ confidence limit. We therefore conclude that there
      are no inconsistencies in our results.
   
    \subsection{Posterior distributions}
\begin{figure}[ht]
	\begin{minipage}[b]{0.33\linewidth}
	  \includegraphics[width=1.0\textwidth]{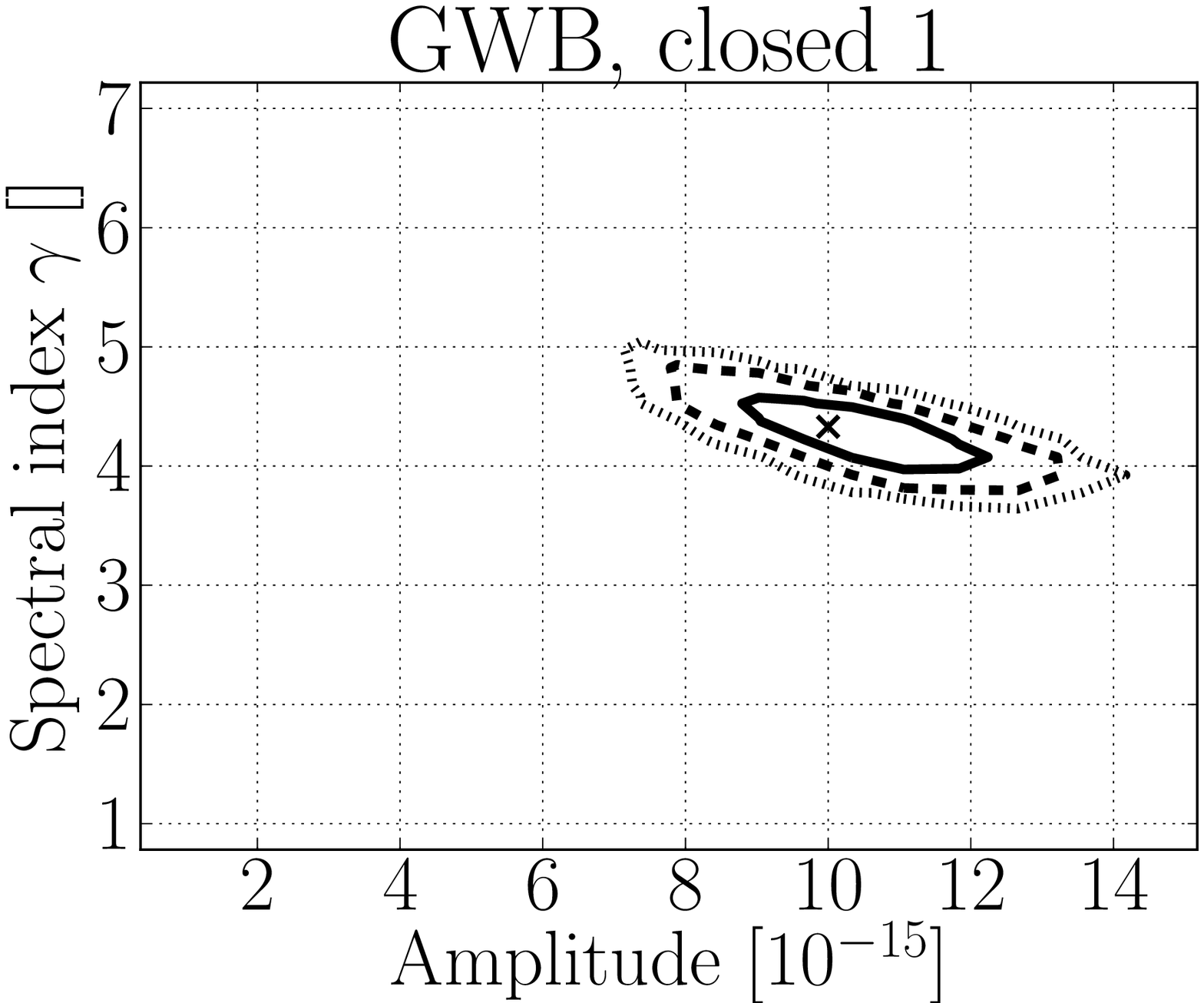}
	  \centering
	\end{minipage}
	\begin{minipage}[b]{0.33\linewidth}
	  \includegraphics[width=1.0\textwidth]{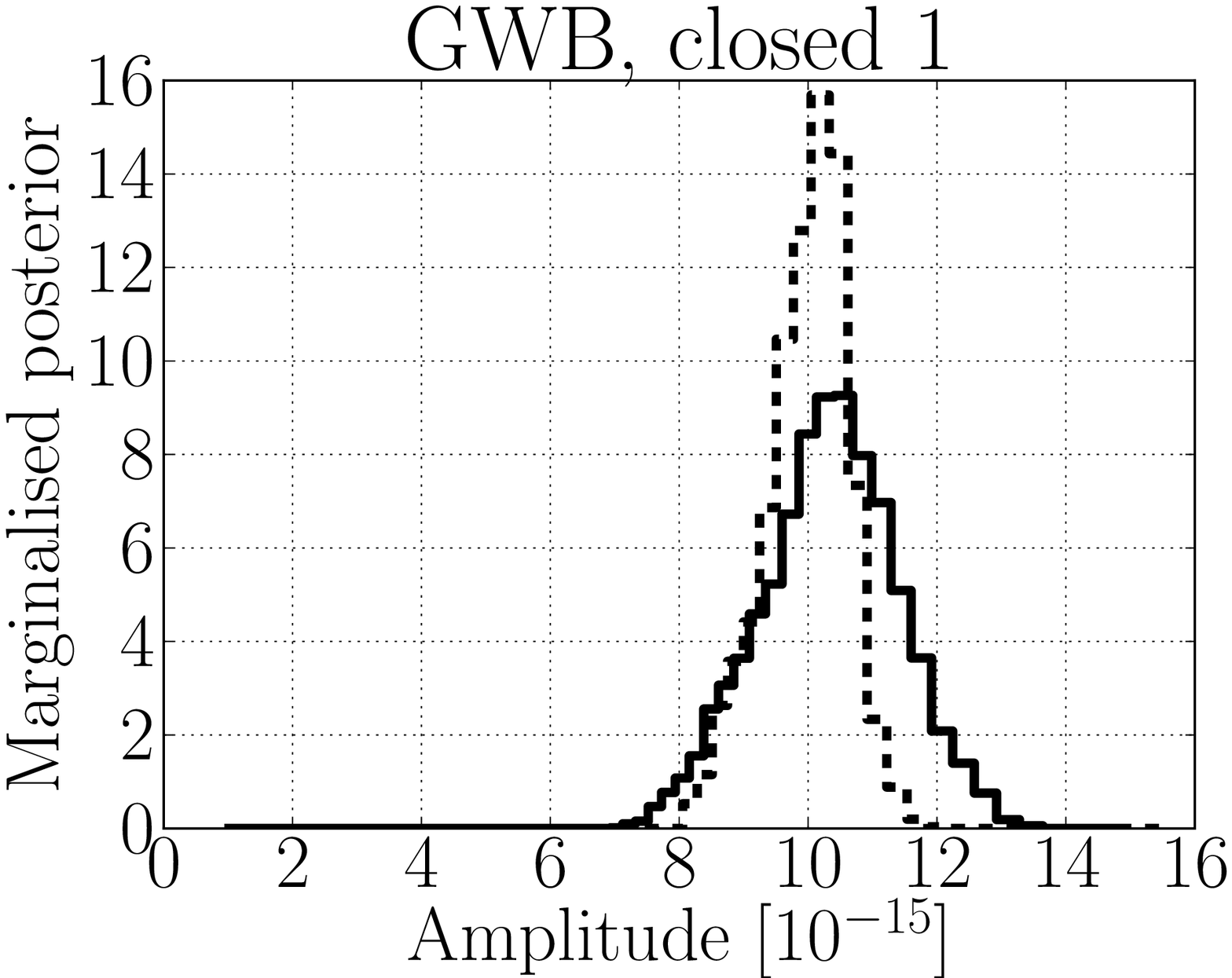}
	  \centering
	\end{minipage}
	\begin{minipage}[b]{0.33\linewidth}
	  \includegraphics[width=1.0\textwidth]{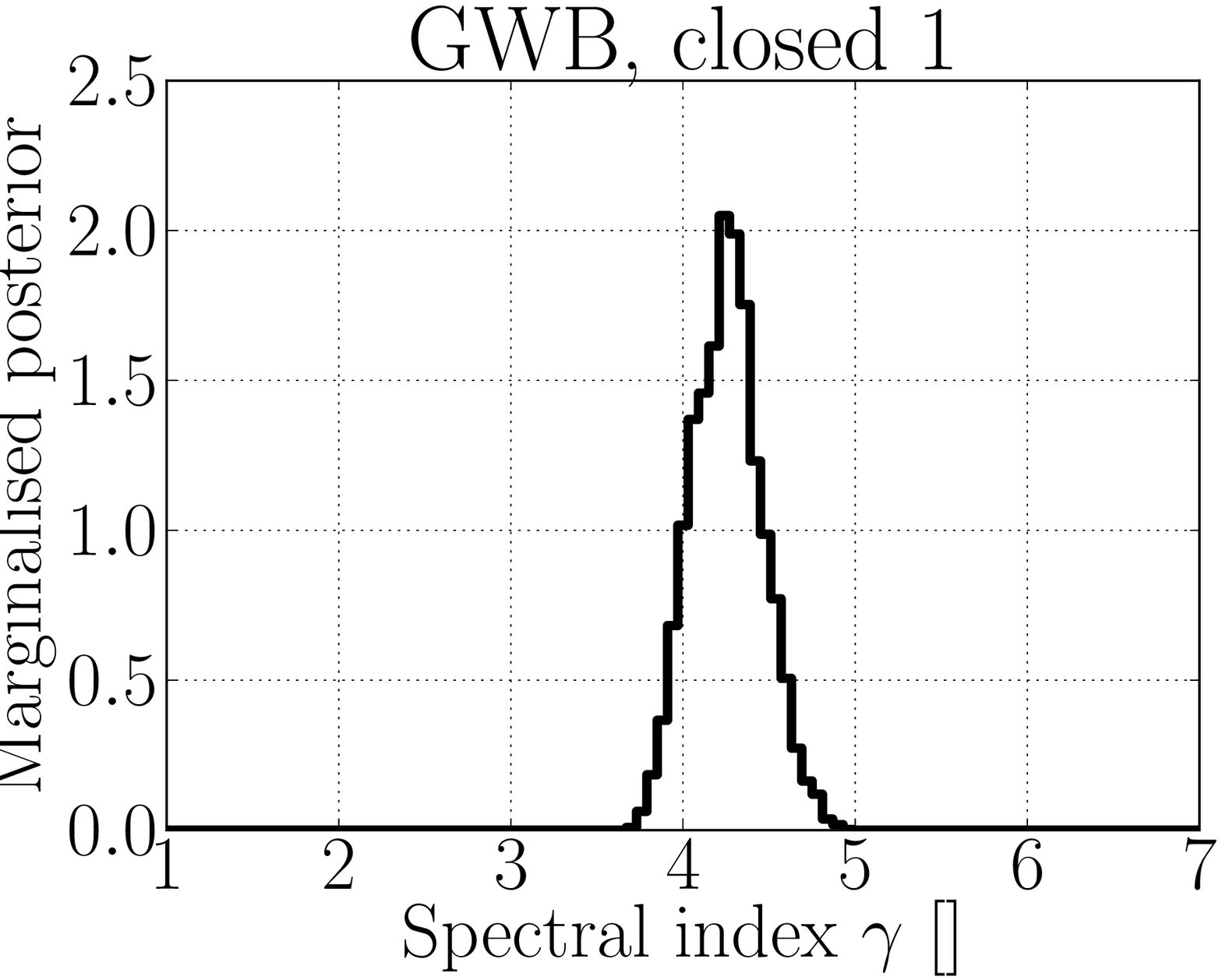}
	  \centering
	\end{minipage}
	\caption{Closed challenge 1, marginalised posterior distributions for a
	model with $36\times 2 + 2 = 74$ free parameters.
	If we run an MCMC with $\gamma_{\rm GWB} = 13/3$ fixed, then we get
	$h_c = 10.0 \pm 0.64 [10^{-15}]$, with a one-sided $2\sigma$ upper-limit of
	$h_c \leq 10.98 [10^{-15}]$}
	\label{fig:closed1}
      \end{figure}

\begin{figure}[ht]
	\label{fig:closed1allnoise}
	\begin{minipage}[b]{0.33\linewidth}
	  \includegraphics[width=1.0\textwidth]{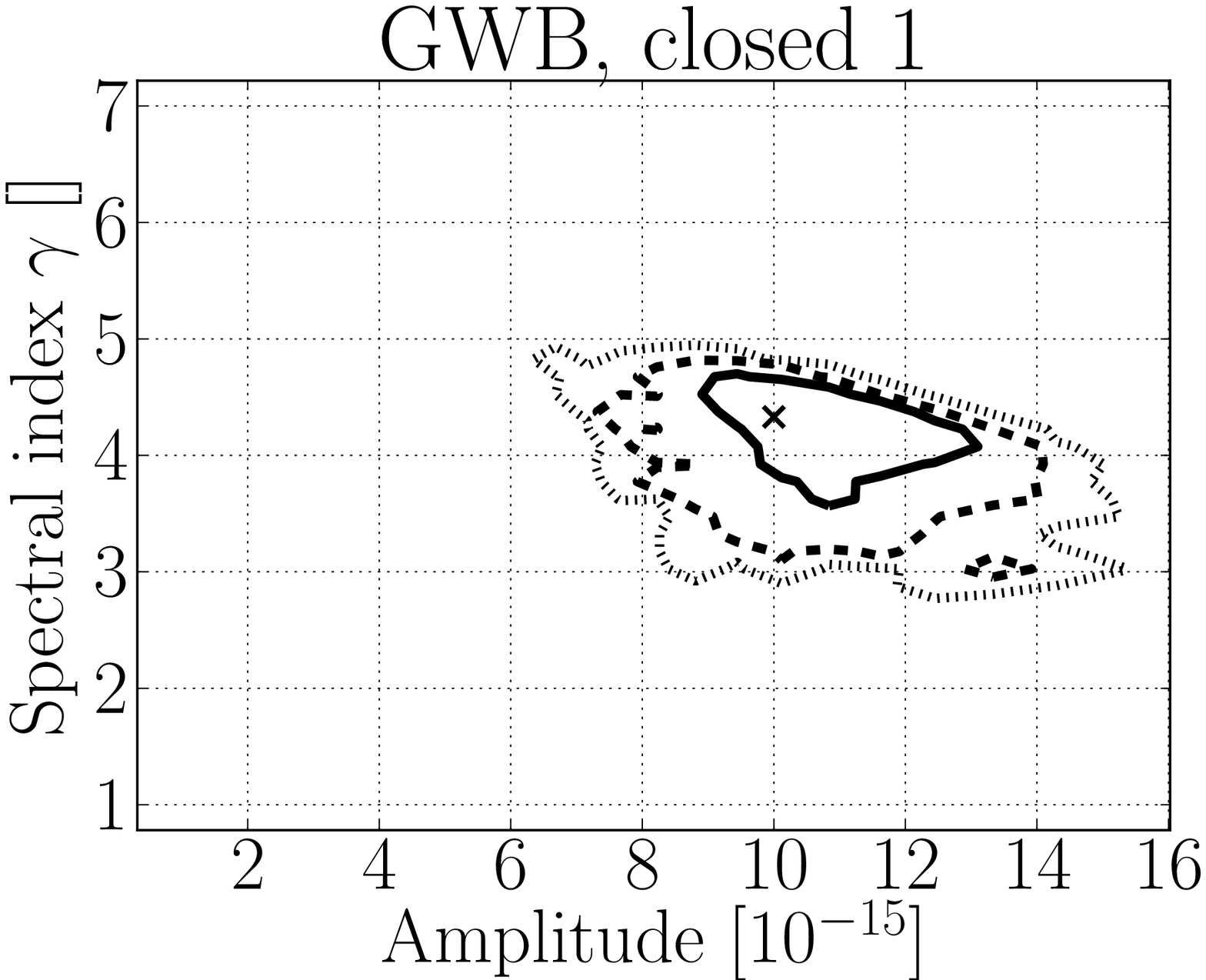}
	  \centering
	\end{minipage}
	\begin{minipage}[b]{0.33\linewidth}
	  \includegraphics[width=1.0\textwidth]{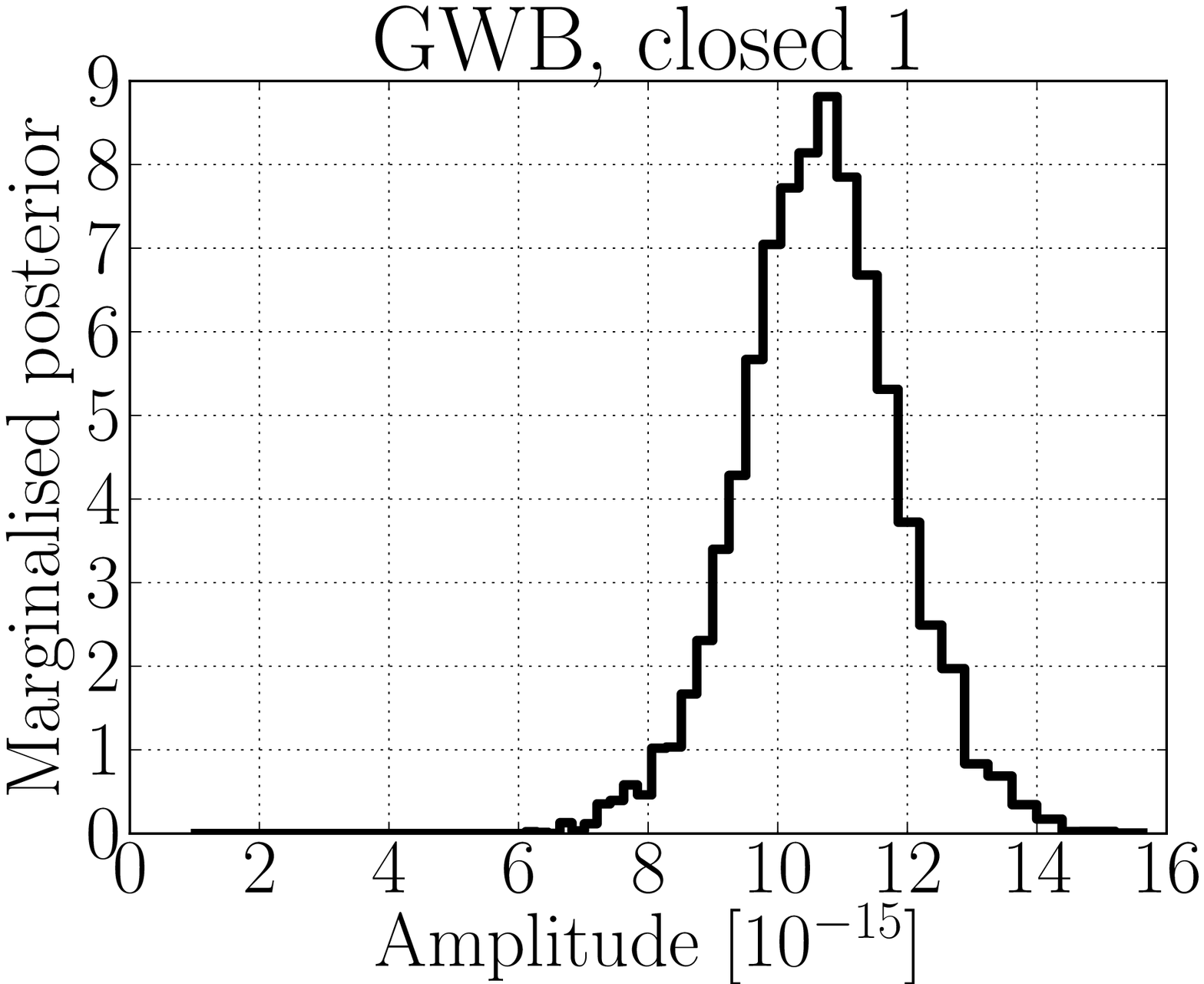}
	  \centering
	\end{minipage}
	\begin{minipage}[b]{0.33\linewidth}
	  \includegraphics[width=1.0\textwidth]{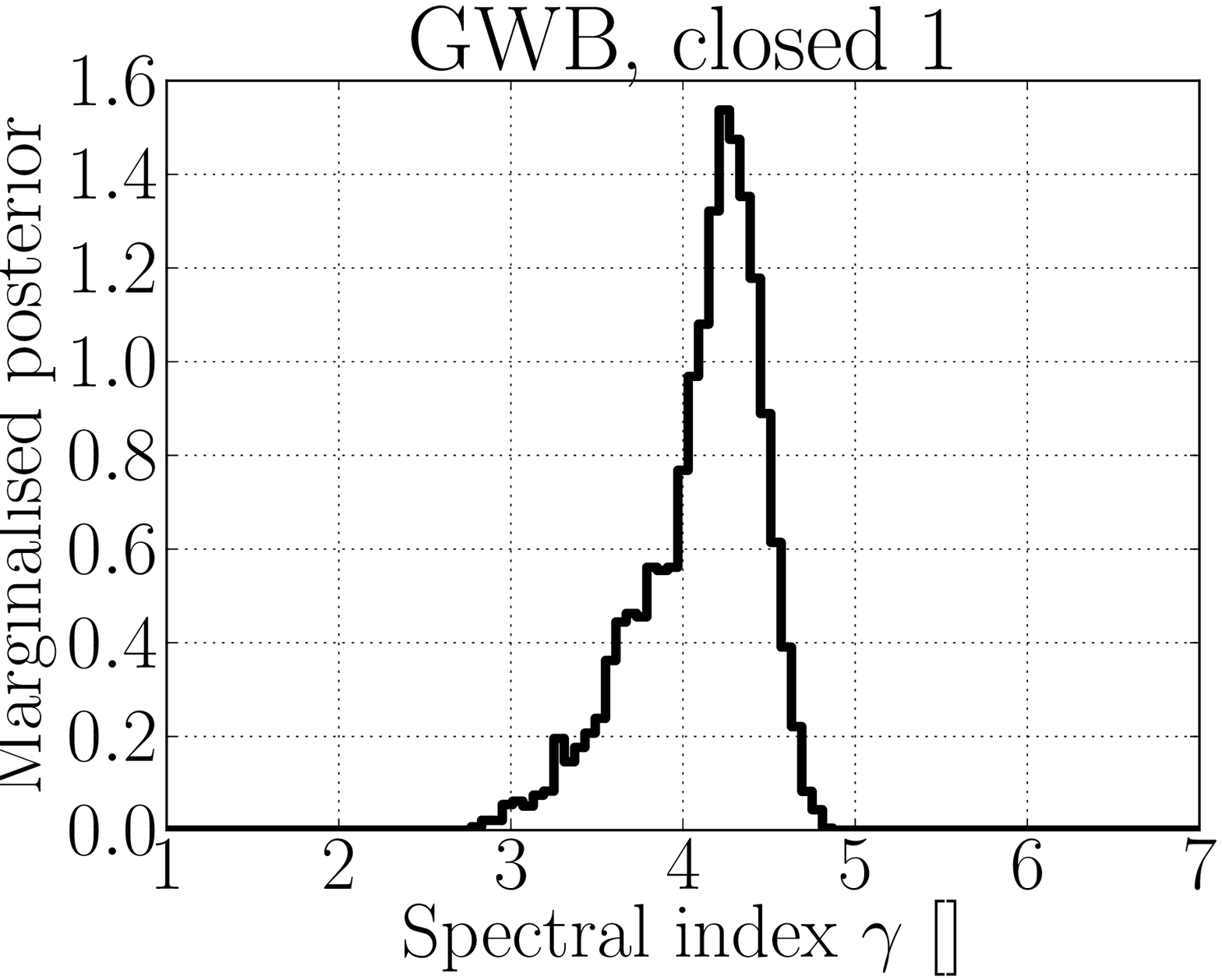}
	  \centering
	\end{minipage}
	\caption{Closed challenge 1, marginalised posterior distributions for a
	model with $4$ free parameters, where all the pulsars are assumed to have an
	equal red noise contribution.}
      \end{figure}

\begin{figure}[ht]
	\label{fig:closed2}
	\begin{minipage}[b]{0.33\linewidth}
	  \includegraphics[width=1.0\textwidth]{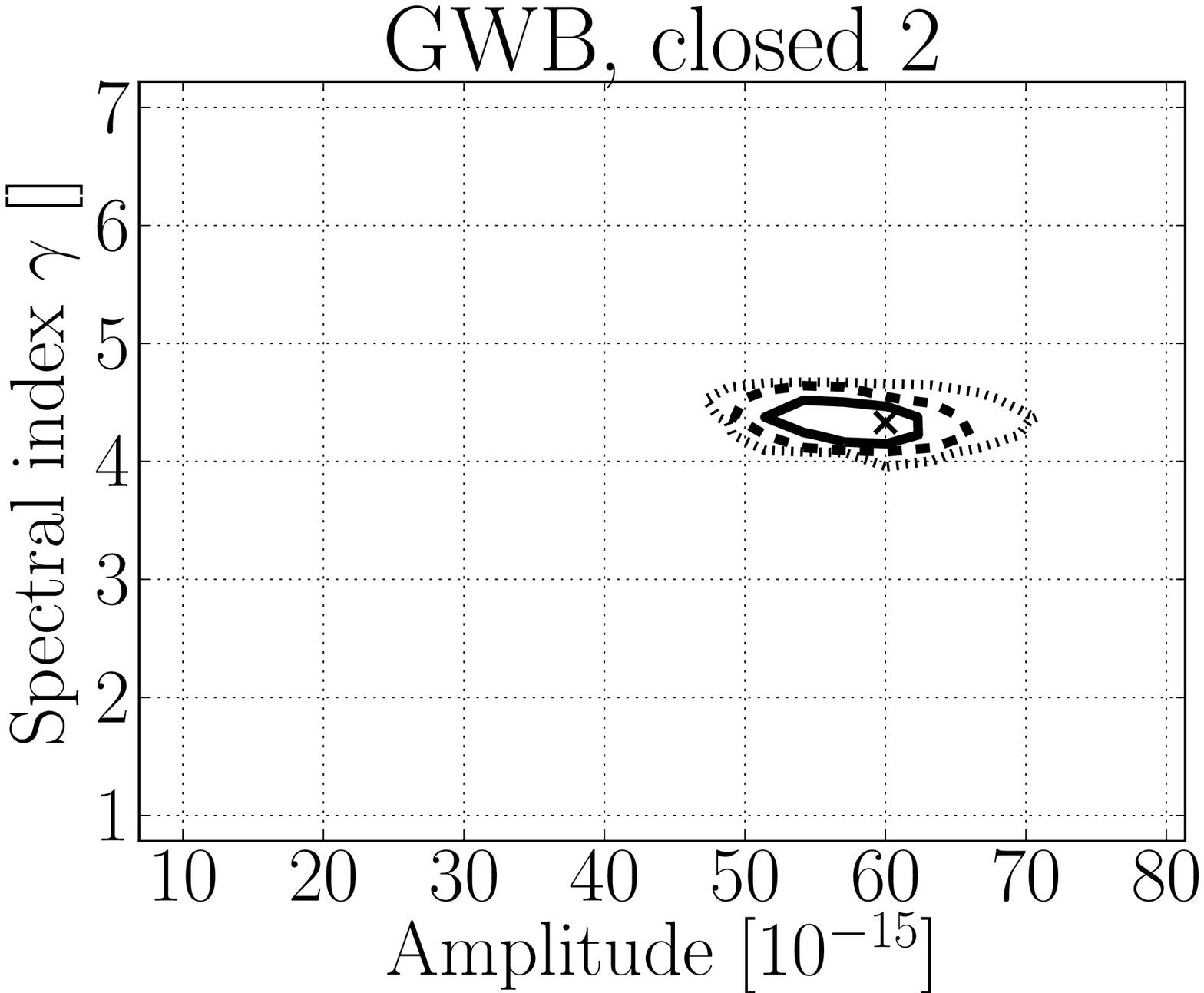}
	  \centering
	\end{minipage}
	\begin{minipage}[b]{0.33\linewidth}
	  \includegraphics[width=1.0\textwidth]{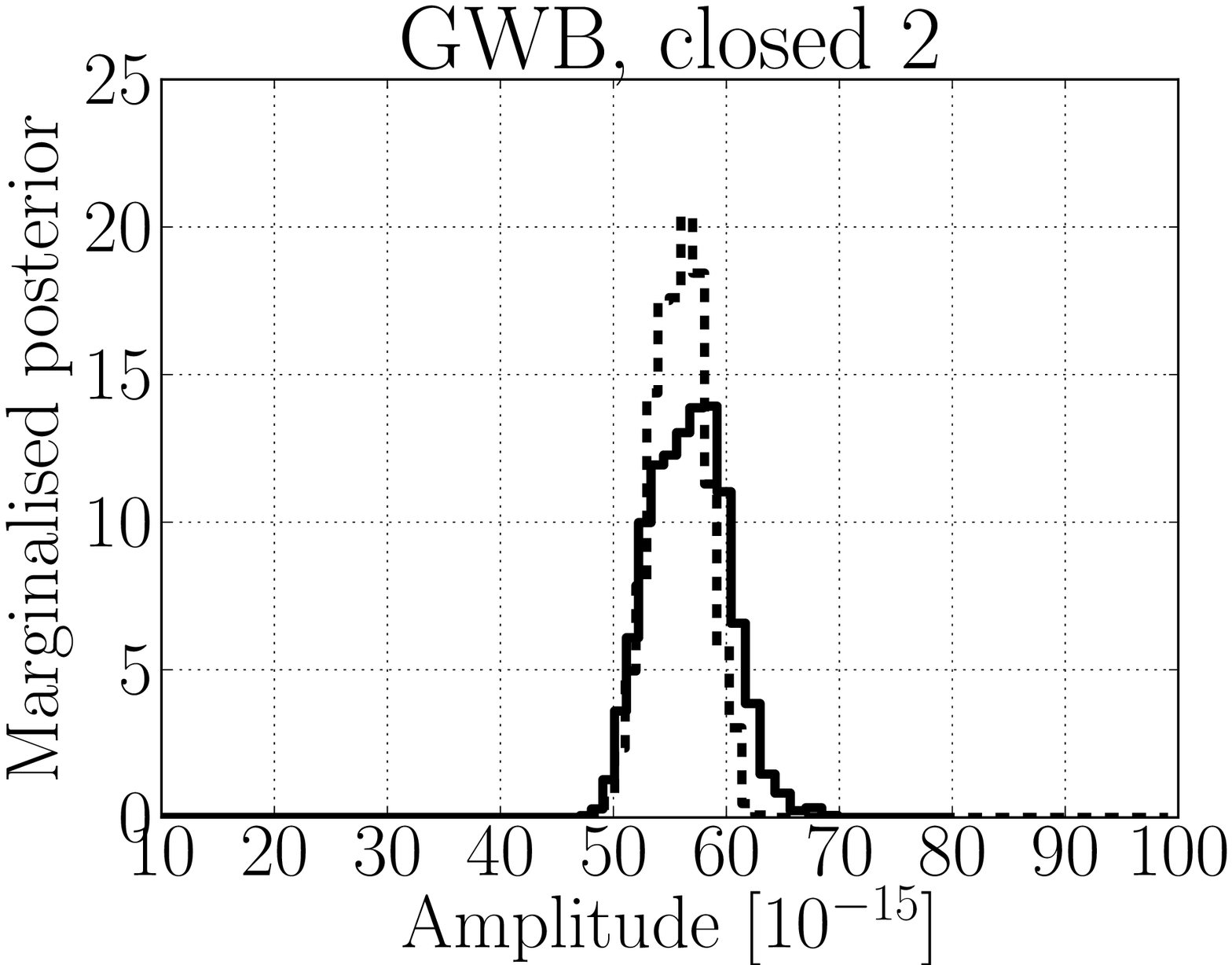}
	  \centering
	\end{minipage}
	\begin{minipage}[b]{0.33\linewidth}
	  \includegraphics[width=1.0\textwidth]{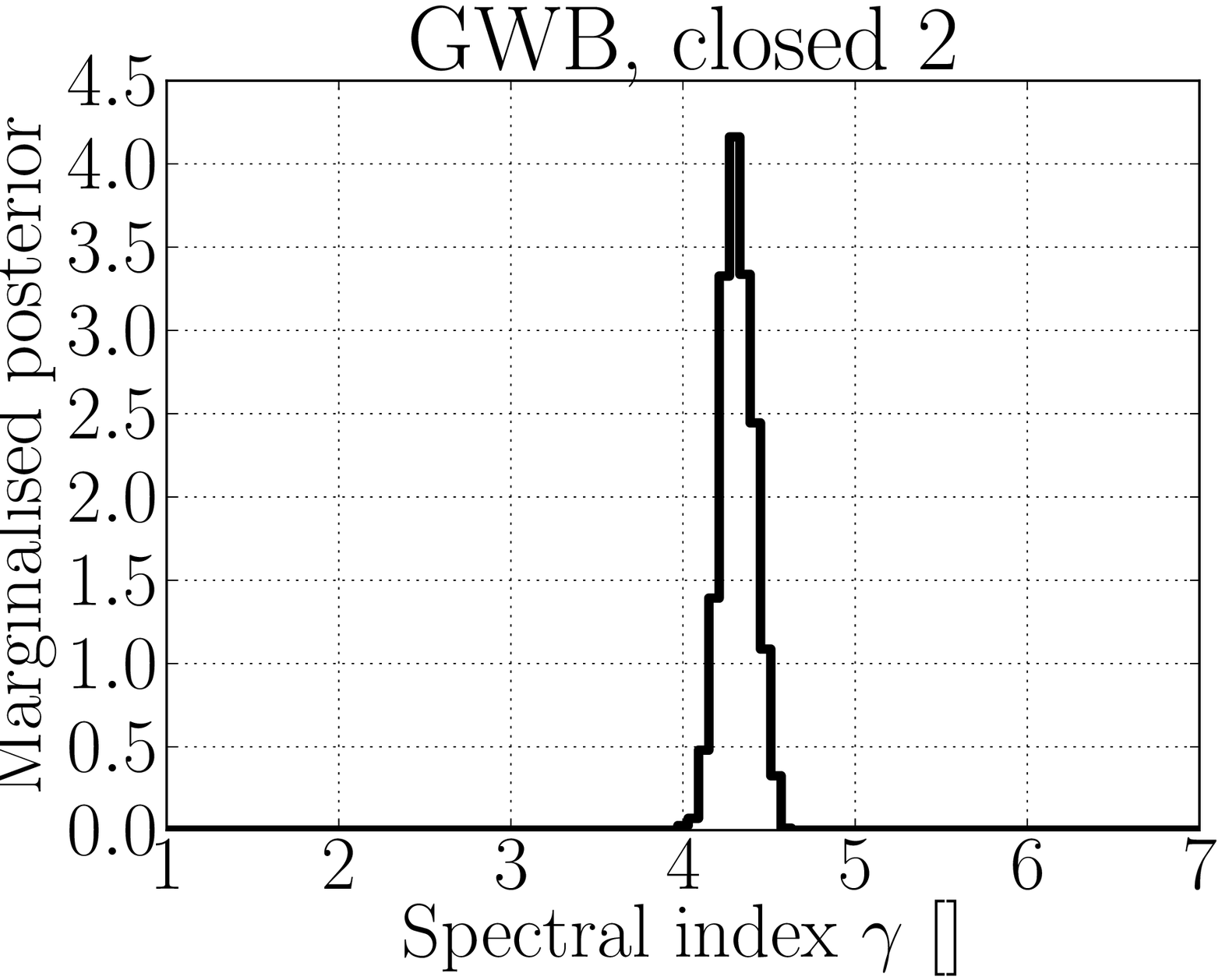}
	  \centering
	\end{minipage}
	\caption{Closed challenge 2, marginalised posterior distributions for a
	model with $36\times 2 + 2 = 74$ free parameters.
	If we run an MCMC with $\gamma_{\rm GWB} = 13/3$ fixed, then we get
	$h_c = 56.3 \pm 2.42 [10^{-15}]$, with a one-sided $2\sigma$ upper-limit of
	$h_c \leq 60.29 [10^{-15}]$}
      \end{figure}

\begin{figure}[ht]
	\begin{minipage}[b]{0.33\linewidth}
	  \includegraphics[width=1.0\textwidth]{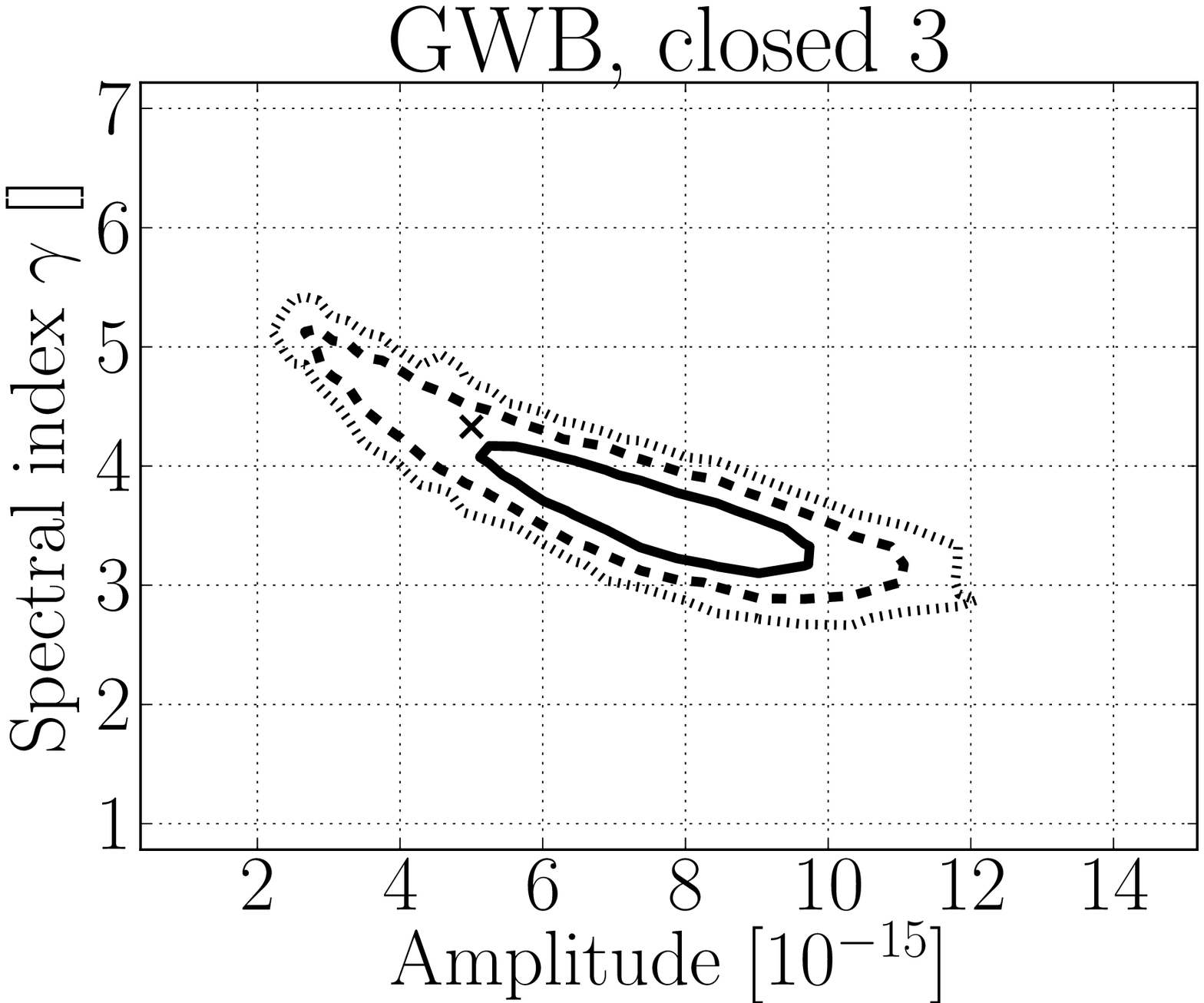}
	  \centering
	\end{minipage}
	\begin{minipage}[b]{0.33\linewidth}
	  \includegraphics[width=1.0\textwidth]{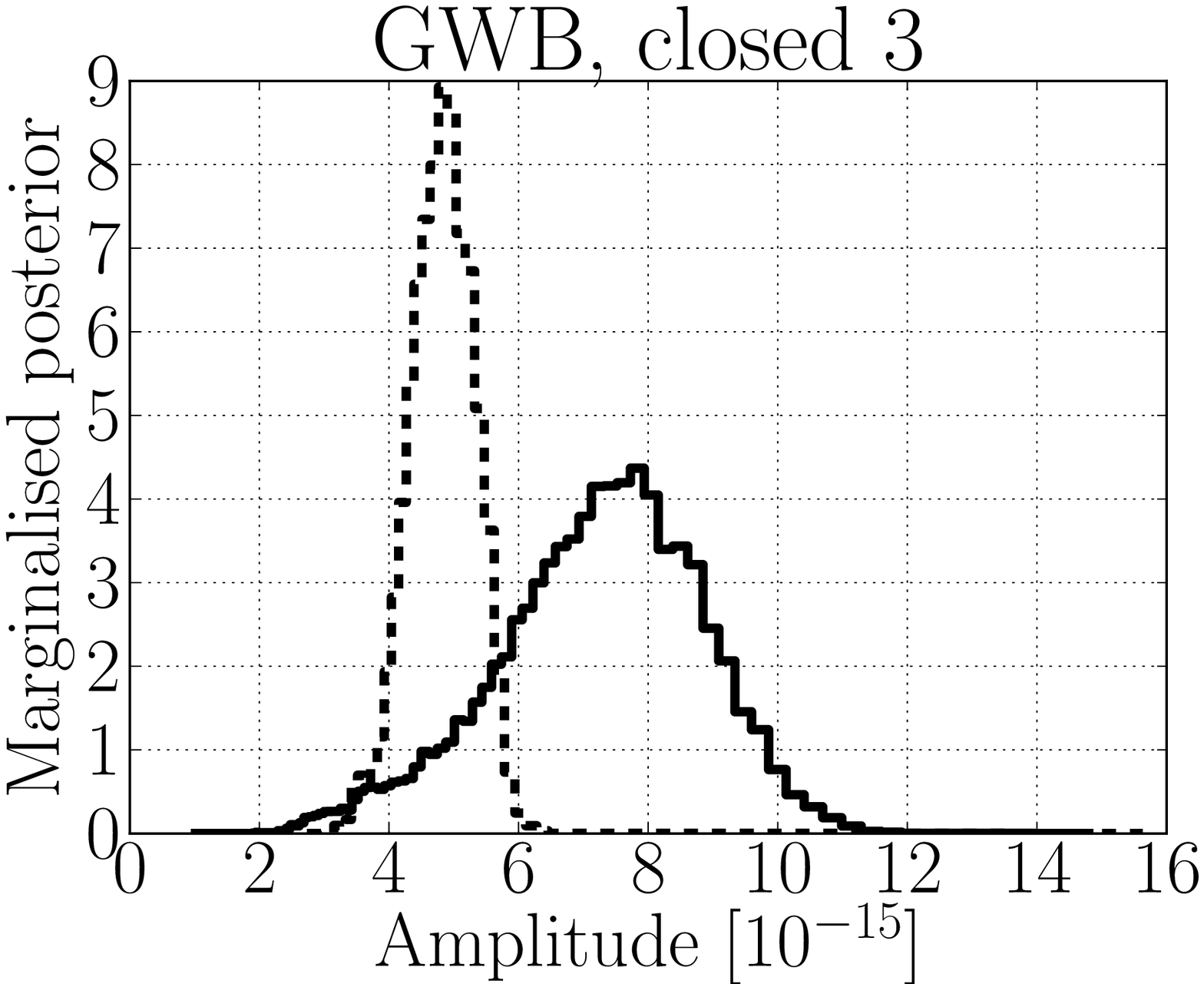}
	  \centering
	\end{minipage}
	\begin{minipage}[b]{0.33\linewidth}
	  \includegraphics[width=1.0\textwidth]{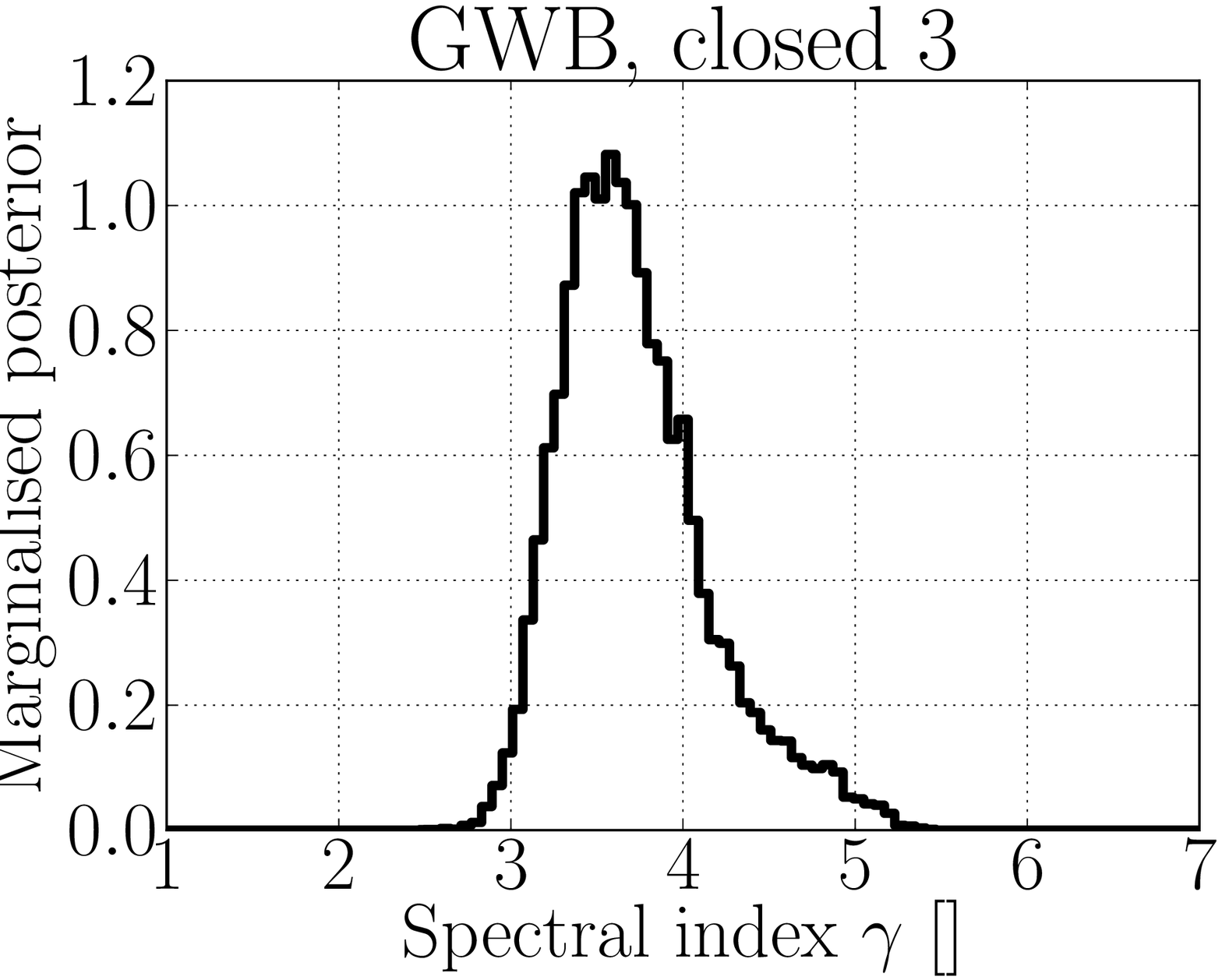}
	  \centering
	\end{minipage}
	\caption{Closed challenge 3, marginalised posterior distributions for a
	model with $36\times 2 + 2 = 74$ free parameters.
	If we run an MCMC with $\gamma_{\rm GWB} = 13/3$ fixed, then we get $h_c
	= 4.83 \pm 0.50 [10^{-15}]$, with a one-sided $2\sigma$ upper-limit of
	$h_c \leq 5.65 [10^{-15}]$}
	\label{fig:closed3}
      \end{figure}

\begin{figure}[ht]
	\begin{minipage}[b]{0.33\linewidth}
	  \includegraphics[width=1.0\textwidth]{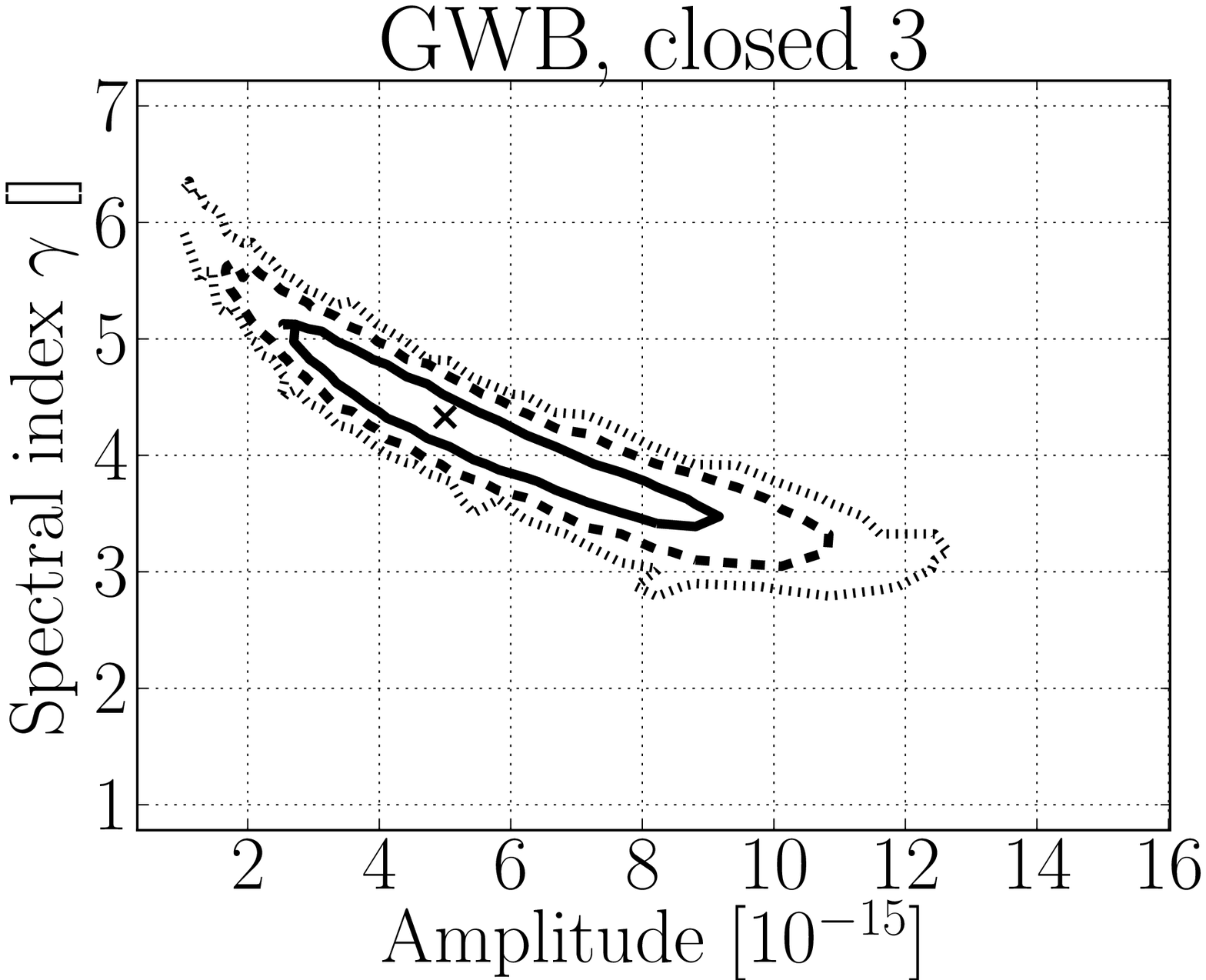}
	  \centering
	\end{minipage}
	\begin{minipage}[b]{0.33\linewidth}
	  \includegraphics[width=1.0\textwidth]{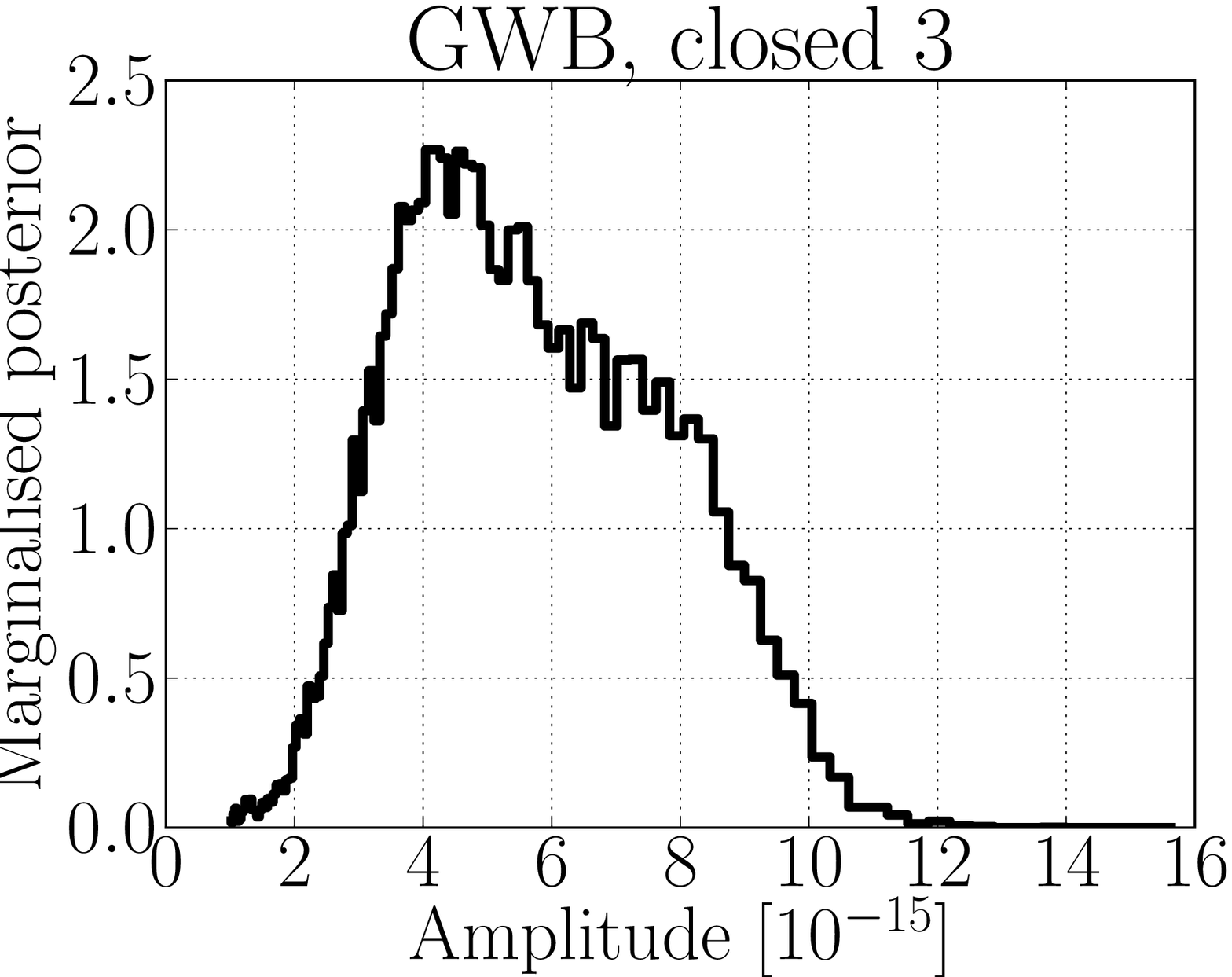}
	  \centering
	\end{minipage}
	\begin{minipage}[b]{0.33\linewidth}
	  \includegraphics[width=1.0\textwidth]{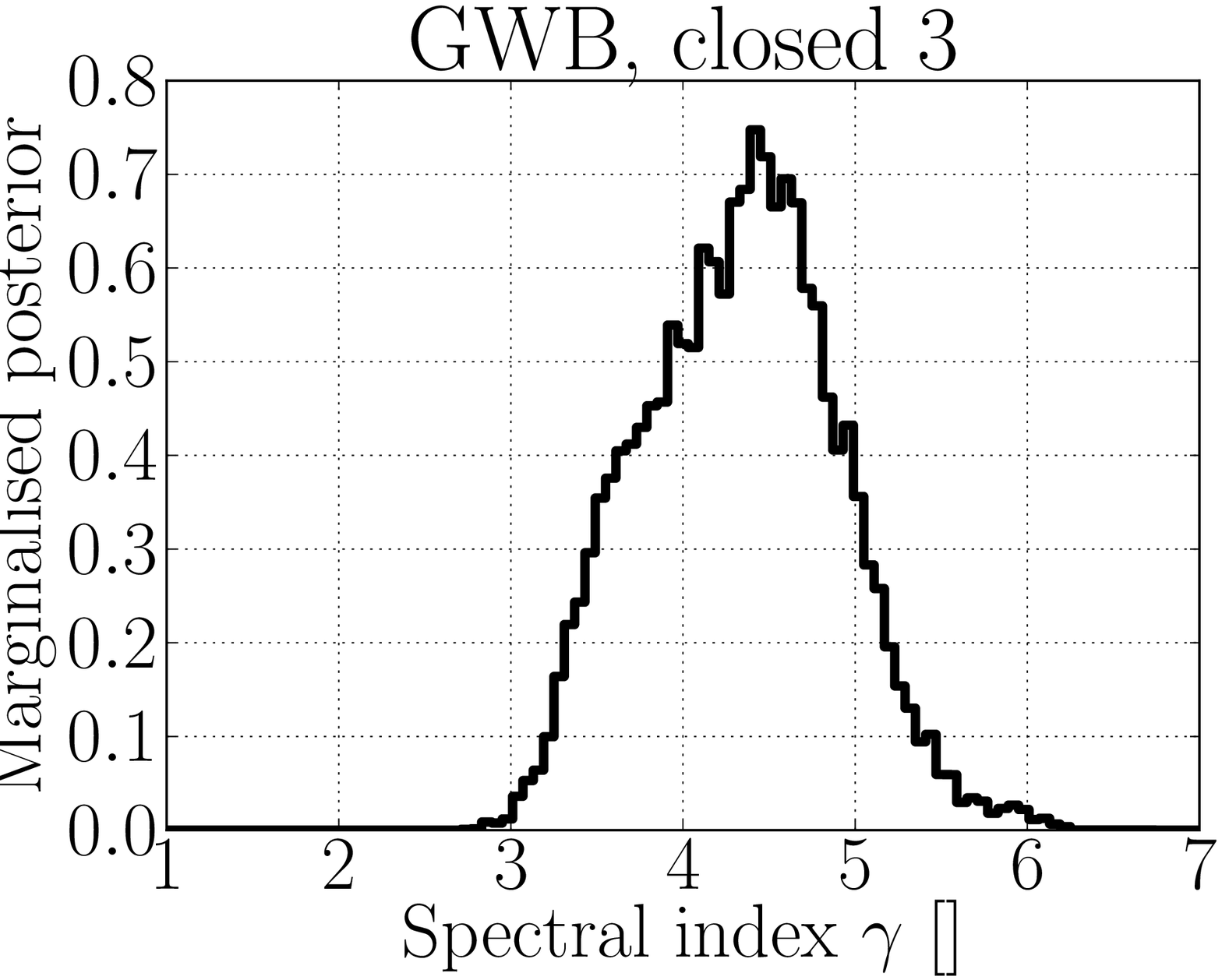}
	  \centering
	\end{minipage}
	\caption{Closed challenge 3, marginalised posterior distributions for a
	model with $4$ free parameters, where all the pulsars are assumed to have an
	equal red noise contribution.}
	\label{fig:closed3allnoise}
      \end{figure}

      In Figure~\ref{fig:closed1}---\ref{fig:closed3allnoise} we present the
      marginalised posterior distribution we found in our analysis.


    \subsection{Computational cost}
      The analysis of the results consisted of four phases: ABC-method
      compression, ABC-method interpolation preparation, twalk burn-in,
      Metropolis-Hastings sampling. For all MDC datasets the computational cost
      was nearly equivalent, with the cost per analysis step:
      \begin{itemize}
	\item
	  $<5$ minutes for ABC-method compression
	\item
	  $<10$ minutes for ABC-method interpolation preparation
	\item
	  $<10$ minutes for the twalk burn-in
	\item
	  $<15$ minutes for the Metropolis-Hastings sampling
      \end{itemize}
      All in all this comes down to almost a full hour on a single machine for
      each dataset. All calculations were done on a MacBook pro, with $2.70$GHz
      CPU, all code linked with a standard LAPACK implementation (Ubuntu
      $12.04$), linked against ATLAS. We did not optimise our code very
      aggressively: instead of using Cholesky-solve, we calculated the inverse
      matrices at various points in the calculations, faster (commercial) LAPACK
      implementations exist, and we did not multi-thread any algorithms.

  \section{Conclusions} \label{sec:conclusions}
    \rvh{Our Bayesian analysis has yielded results consistent with the injected
      values of the first IPTA Mock Data Challenge.  Our analysis indicates the
      presence of a stochastic background of gravitational waves in all three
      data challenge sets, with amplitudes of: $h_c = 1.03 \pm 0.11 [\times
      10^{-14}]$, $h_c = 5.70 \pm 0.35 [\times 10^{-14}]$, and $h_c = 6.91 \pm
      1.72 [\times 10^{-15}]$. We found for the spectral index for the
      gravitational wave signal: $\gamma = 4.28 \pm 0.20$, $\gamma = 4.35 \pm
      0.09$, and $\gamma = 3.75 \pm 0.40$, with a strong covariance not
      reflected in these numbers between the signal and the red noise for closed
      challenge three. If we constrain the GWB spectral index to the usual
      choice of $\gamma = 13/3$, we obtain the estimates: $h_c = 10.0 \pm 0.64
      [10^{-15}]$, $h_c = 56.3 \pm 2.42 [10^{-15}]$, and $h_c = 4.83 \pm 0.50
      [10^{-15}]$, with one-sided $2\sigma$ upper-limits of: $h_c \leq 10.98
      [10^{-15}]$, $h_c \leq 60.29 [10^{-15}]$, and $h_c \leq 5.65 [10^{-15}]$.}
      
    \rvh{Tests on the open
      challenges have indicated that the prior for the noise and GWB amplitudes
      needs to be uniform on a logarithmic scale. 
      Future work will include more tests of convergence of the MCMC chain,
      internal consistency checks.}

  \section*{Acknowledgements}
    We would like to thank the organisers of the Mock Data Challenge for their
    work in setting up this challenge: Fredrick A. Jenet, Michael Keith, and
    Kejia Lee. The data challenge has brought some potential issues to our
    attention that require further study.


  \bibliographystyle{authordate1}
  \bibliography{mdc1-epta}

\begin{thebibliography}{}

\bibitem[\protect\citename{{Christen} \& {Fox}, }2010]{Christen2010}
{Christen}, J.~A., \& {Fox}, C. 2010.
\newblock A general purpose sampling algorithm for continuous distributions
  (the t-walk).
\newblock {\em Bayesian Anal.}, {\bf 5}(Nov.), 263--281.

\bibitem[\protect\citename{{Edwards} {\it et~al.}, }2006]{Edwards2006}
{Edwards}, R.~T., {Hobbs}, G.~B., \& {Manchester}, R.~N. 2006.
\newblock {TEMPO2, a new pulsar timing package - II. The timing model and
  precision estimates}.
\newblock {\em \mnras}, {\bf 372}(Nov.), 1549--1574.

\bibitem[\protect\citename{{Feroz} {\it et~al.}, }2009]{Feroz2009}
{Feroz}, F., {Hobson}, M.~P., \& {Bridges}, M. 2009.
\newblock {MULTINEST: an efficient and robust Bayesian inference tool for
  cosmology and particle physics}.
\newblock {\em \mnras}, {\bf 398}(Oct.), 1601--1614.

\bibitem[\protect\citename{{Foreman-Mackey} {\it et~al.}, }2012]{Foreman2012}
{Foreman-Mackey}, D., {Hogg}, D.~W., {Lang}, D., \& {Goodman}, J. 2012.
\newblock {emcee: The MCMC Hammer}.
\newblock {\em ArXiv e-prints}, Feb.

\bibitem[\protect\citename{Hellings \& Downs, }1983]{Hellings1983}
Hellings, R.W., \& Downs, G.S. 1983.
\newblock Upper limits on the isotropic gravitational radiation background from
  pulsar timing analysis.
\newblock {\em \apj}, {\bf 265}, L39--L42.

\bibitem[\protect\citename{{Hobbs} {\it et~al.}, }2006]{Hobbs2006}
{Hobbs}, G.~B., {Edwards}, R.~T., \& {Manchester}, R.~N. 2006.
\newblock {TEMPO2, a new pulsar-timing package - I. An overview}.
\newblock {\em \mnras}, {\bf 369}(June), 655--672.

\bibitem[\protect\citename{Press {\it et~al.}, }1992]{Press1992}
Press, William, Teukolsky, Saul, Vetterling, William, \& Flannery, Brian. 1992.
\newblock {\em {Numerical Recipes in C}}. 2nd edn.
\newblock Cambridge, UK: Cambridge University Press.

\bibitem[\protect\citename{{Roberts} {\it et~al.}, }1997]{Roberts1997}
{Roberts}, G.O., {Gelman}, A., \& {Gilks}, W.R. 1997.
\newblock {Weak convergence and optimal scaling of random walk Metropolis
  algorithms}.
\newblock {\em Ann. of Appl. Prob.}, {\bf 7}(1), 110--120.

\bibitem[\protect\citename{{Taylor} {\it et~al.}, }2012]{Taylor2012}
{Taylor}, S.~R., {Gair}, J.~R., \& {Lentati}, L. 2012.
\newblock {Weighing The Evidence For A Gravitational-Wave Background In The
  First International Pulsar Timing Array Data Challenge}.
\newblock {\em ArXiv e-prints}, Oct.

\bibitem[\protect\citename{{van Haasteren}, }2012]{vanhaasteren2012b}
{van Haasteren}, R. 2012.
\newblock {Accelerating pulsar timing data analysis}.
\newblock {\em ArXiv e-prints}, Oct.

\bibitem[\protect\citename{{van Haasteren} \& {Levin}, }2010]{vanhaasteren2010}
{van Haasteren}, R., \& {Levin}, Y. 2010.
\newblock {Gravitational-wave memory and pulsar timing arrays}.
\newblock {\em \mnras}, {\bf 401}(Feb.), 2372--2378.

\bibitem[\protect\citename{{van Haasteren} \& {Levin},
  }2012]{vanhaasteren2012a}
{van Haasteren}, R., \& {Levin}, Y. 2012.
\newblock {Understanding and analysing time-correlated stochastic signals in
  pulsar timing}.
\newblock {\em ArXiv e-prints}, Feb.

\bibitem[\protect\citename{{van Haasteren} {\it et~al.},
  }2009]{vanhaasteren2009}
{van Haasteren}, R., {Levin}, Y., {McDonald}, P., \& {Lu}, T. 2009.
\newblock {On measuring the gravitational-wave background using Pulsar Timing
  Arrays}.
\newblock {\em \mnras}, {\bf 395}(May), 1005--1014.

\bibitem[\protect\citename{{van Haasteren} {\it et~al.},
  }2011]{vanhaasteren2011}
{van Haasteren}, R., {Levin}, Y., {Janssen}, G.~H., {Lazaridis}, K., {Kramer},
  M., {Stappers}, B.~W., {Desvignes}, G., {Purver}, M.~B., \& {Lyne}, A.~G.
  2011.
\newblock {Placing limits on the stochastic gravitational-wave background using
  European Pulsar Timing Array data}.
\newblock {\em \mnras}, {\bf 414}(July), 3117--3128.

\end{thebibliography}

\end{document}